\documentclass[12pt]{article}

\usepackage{amsmath,amsthm,amsfonts, latexsym}
\usepackage{graphicx}

\newcommand{\mtx}[2]{\left(\begin{array}{#1}#2\end{array}\right)}

\begin{document}

\begin{center}

\bigskip
{\Large Picturing qubits in phase space}\\

\bigskip

William K.~Wootters\\

\bigskip

{\small{\sl

Department of Physics, Williams College, Williamstown, 
MA 01267, USA }}\vspace{3cm}

\end{center}
\subsection*{\centering Abstract}
{Focusing particularly on one-qubit and two-qubit
systems, I explain how the quantum state of a system of $n$ qubits
can be expressed as a real function---a generalized Wigner
function---on a discrete $2^n \times 2^n$ phase space.  The 
phase space is based on the finite field having
$2^n$ elements, and its geometric structure leads 
naturally to the construction of a complete set
of $2^n+1$ mutually conjugate bases.}

\vfill

PACS numbers: 03.65.Ca, 03.65.Ta, 03.65.Wj, 02.10.De

\newpage

On Charlie Bennett's main webpage one finds two photographs:
one of Charlie himself and the other of a vortex created
by a beaver dam.  A vortex is a wonderful example of a structure
that maintains its form by not holding on to its substance; it
thrives because it continually gives its material away. 

Some summers ago I was supervising four undergraduates
in research projects in quantum information theory, and together
we
drove down to the Watson Research Center for a day to 
talk with Charlie.  He took us
to the Croton dam, one of his favorite places.  
As we sat there on
the dam with the sound of water in the background, we 
discussed quantum information and wrote
down quantum states 
on a large pad of newsprint that Charlie had
brought along.  The breeze was blowing a bit, and as always around 
Charlie, ideas
were swirling. We talked for hours.  
In the years that have passed 
since that afternoon, traces of that experience
and traces of those ideas 
have surely been carried far out into the world---who
knows how far---in the
lives of those four students and the people they have encountered.  
I offer this little story as one example of hundreds of similar 
acts of sharing, through which Charlie
Bennett has had an influence on the world of science
that could never be captured by any 
reckoning based on cited publications.  
It is a pleasure to dedicate this paper to him on
the occasion of his sixtieth birthday.    

A spin-1/2 particle is probably not a vortex, though
there may be some virtue in thinking of it more as 
a process than a static object.  But here I will
not be adventurous in that way.  This paper is about
qubits as normally conceived, 
and I will use the spin of a spin-1/2 particle
as my standard physical example of a qubit.  We usually express
the quantum state of a system of qubits as a state vector or
density matrix.  The main point of this paper is to show how
one can represent such a quantum state as a real function
on a {\em phase space}, not a continuous phase space whose
axes stand for position and momentum, but a discrete phase
space whose axes are associated with a pair of conjugate bases
for the finite-dimensional state space.  
Much of the work I report here
was done jointly with Kathleen Gibbons, and many of the mathematical
details are 
given in Ref.~\cite{Gibbons}.  Here I want to lay out the
overall contours of this phase space construction.   

Discrete phase space representations
have been proposed in a number of
earlier papers [2--13].
The particular representation to be 
described here is 
different in ways that I will discuss later.  First, however,
I would like to motivate the work by posing what might
seem to be an unrelated problem, the problem of determining
an unknown quantum state.  

\section{State determination} \label{state}

Imagine a device whose output is a beam of spin-1/2
particles.  We do not know enough about the device to
predict the spin state of the particles that it
produces.  They might all be in the spin-up state, for example, or
they might be completely unpolarized.  We do, however,
assume that the device does not change its operation
significantly over time, so that we ought to be able
to describe the whole ensemble of particles by a single
state (possibly a mixed state) of a single spin-1/2 particle.  
A general spin state of a spin-1/2 particle can be 
pictured as a point either on the surface or in the
interior of a unit sphere.  Points on the surface of
the sphere
are pure states, points in the interior are
mixed states, and the center of the sphere is
the completely mixed or completely unpolarized state.
Our job is to perform a set of measurements on the
particles so as to determine which point represents
the spin state actually produced by this mysterious device.  

How do we proceed?  Suppose we perform the measurement
``up vs down'' on the first hundred particles that
come our way.  What do we get from these measurements?   We
get a rough estimate of the 
vertical height of the point that represents the ensemble's
state, because the height is what
determines the probabilities of ``up'' and ``down''.  
However we do not get a perfectly precise value of the 
height, because we have performed only a finite number
of trials and therefore still have some statistical error.
We know that we 
will have to live with some statistical error, so we 
now turn our attention to pinning the state down better
along the horizontal
dimensions.  
In order to do this, we perform measurements of spin
along two other axes.  (Of course we have to use new 
particles for these measurements.  The ones we have 
already measured hold no further information for us.)
If we call the vertical direction $z$,
we might let our two new measurements be measurements
of spin along the $x$ and $y$ directions.  In this way we can narrow 
the range of likely states to a small region, typically in the
interior of the
sphere.  

\begin{figure}[h]
\centering
\includegraphics{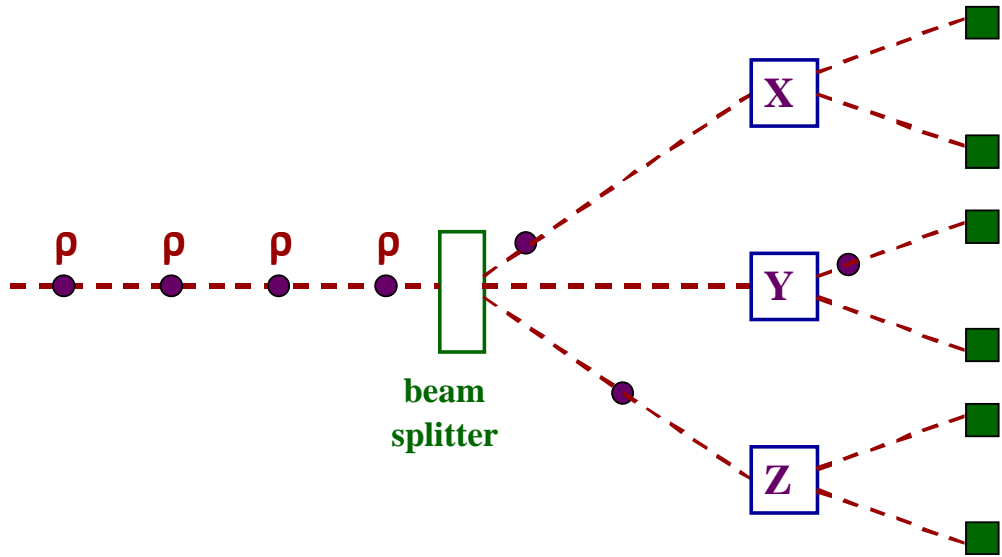}  
\caption{Determining the state of a single qubit.  The
beam splitter represents the random splitting of the original
ensemble into three subensembles.}
\label{3meas}
\end{figure}

It is clear that if we restrict our attention to orthogonal
measurements, each represented by a pair of 
diametrically opposite directions
in space, then in order to have any hope of pinning 
down the state we need to use at least three different
measurements.  That is, we need to break the whole
ensemble into at least three subensembles and measure
each of these subensembles along a different axis.  
As long as the three axes are not coplanar, we will
eventually get an arbitrarily good estimate of the
state.  But some non-coplanar choices are better than
others: the statistical error will be 
minimized if we use three axes that are perpendicular
to each other, such as the 
$x$, $y$, and $z$ axes as imagined above \cite{Fields}.  In this
case the three measurements are called ``mutually
conjugate,'' meaning that each eigenvector of one
measurement is an {\em equal} superposition 
of the eigenvectors
of any of the other measurements.\footnote{Normally
I call such measurements ``mutually unbiased,'' but
as a tribute to Charlie on this occasion, I use the
nomenclature that he prefers.}  If we choose the 
measurements in this way, each different
measurement gives us information that is as independent
as possible
of the information provided by the other measurements.
A state-determination
scheme based on measurements in the $x$, $y$, and $z$
directions is pictured symbolically in Fig.~\ref{3meas},
where the measurements are labeled $X$, $Y$, and $Z$.\footnote{In 
this discussion I am assuming that each qubit is measured
independently and without reference to the results obtained
from other qubits.  A more efficient approach---in the 
sense of reducing the number of qubits needed---would be to 
use an adaptive scheme \cite{Fischer} or a holistic
measurement of all the qubits together \cite{Derka}, but
I do not consider these more sophisticated strategies here.  For
a review of the problem of quantum state reconstruction, see
Ref.~\cite{Buzek}.}

\begin{figure}[h]
\centering
\includegraphics{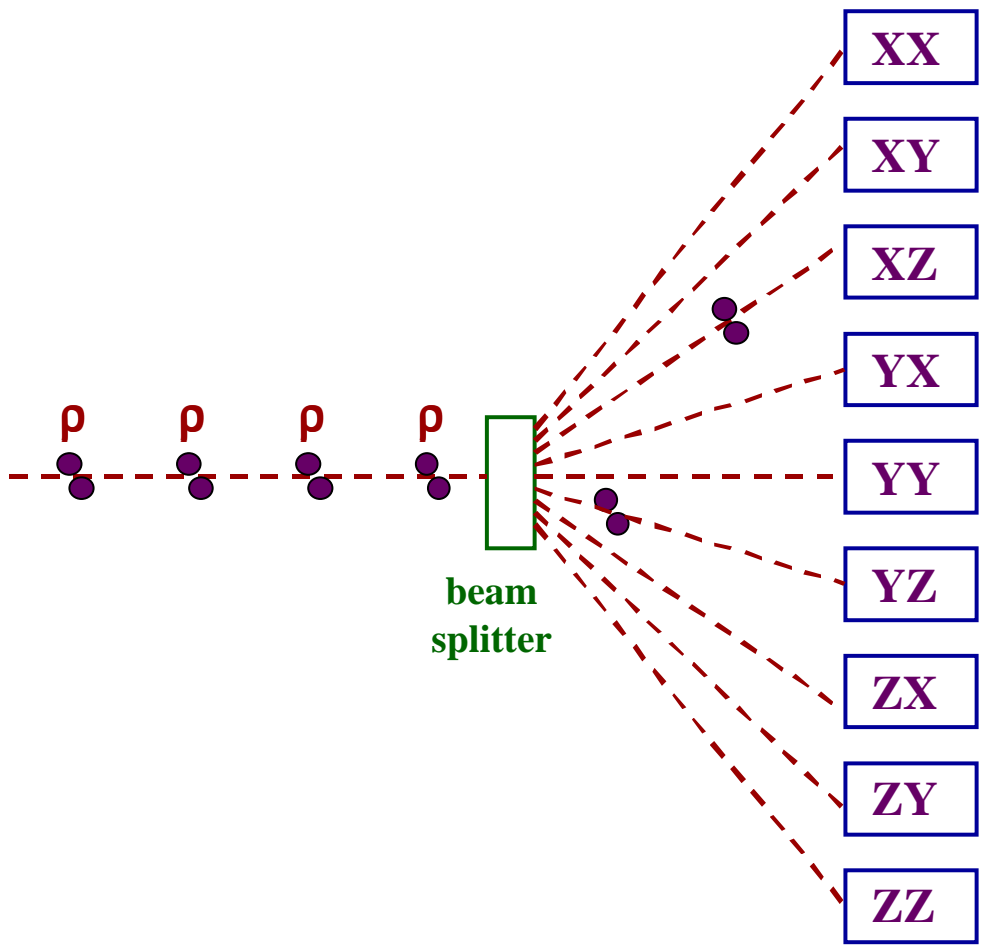}  
\caption{Determining the state of a pair of qubits.}
\label{9meas}
\end{figure}

Let us now consider the problem of state determination
for a {\em pair} of qubits.  We imagine a device that
produces a beam of pairs of spin-1/2 particles.  For example,
each pair might consist of two distinguishable
spin-1/2 nuclei in 
the same molecule.  How might we go about determining
the state of one of these pairs?  One can show that
it is sufficient to use the following nine measurements,
each performed on a different subensemble \cite{Wootters2}:
$XX$, $XY$, $XZ$, $YX$, $YY$, $YZ$, $ZX$, $ZY$, $ZZ$.
Here $XY$, for example, means measuring the first particle
of the pair along the $x$ axis and the second along the
$y$ axis.  This scheme is illustrated in Fig.~\ref{9meas}.

In a certain sense, this nine-measurement scheme is not as 
efficient as it might be.  A general density matrix
for a pair of qubits requires $4^2 - 1 = 15$  real
parameters for its specification.  A general orthogonal
measurement with four outcomes (such as any of the nine
measurements listed above) provides $4-1=3$ independent
probabilities, since the probabilities must sum to unity.
Therefore, if we restrict our attention to orthogonal
measurements, we need at least $15/3 = 5$ such measurements
to determine the state.  The above scheme thus uses more
distinct measurements than would seem to be necessary.
We should be able to get away with using only five
distinct measurements, and ideally these five measurements
would be mutually conjugate so as to minimize the
statistical error.\footnote{This notion of ``efficient,''
{\em i.e.}, using as few distinct measurements
as possible, is perhaps a bit artificial, though one
can imagine that there may be some experimental advantage
in achieving this sort of efficiency.  If the
issue is only to minimize the number of qubits 
measured, then one can do just as well with a very 
large number of distinct measurements \cite{Maccone}.}

We are thus led to an interesting question:
do there exist five orthogonal measurements for a pair
of qubits that are all mutually conjugate?  The answer is
not at all obvious, but it is known: yes,
there do exist such measurements 
\cite{Fields,Calderbank,Bandy,Lawrence}.  
One set that satisfies
all the imposed constraints is the following: $XX$, $YY$,
$ZZ$, plus two other measurements that are both closely
related to the Bell measurement.  The Bell measurement,
named for John Bell, has the following eigenstates:
\begin{eqnarray}
(1/\sqrt{2})(\mid \uparrow \uparrow \rangle \nonumber
+\mid \downarrow \downarrow \rangle ) \\
(1/\sqrt{2})(\mid \uparrow \uparrow \rangle \nonumber
-\mid \downarrow \downarrow \rangle ) \\
(1/\sqrt{2})(\mid \uparrow \downarrow \rangle \nonumber
+\mid \downarrow \uparrow \rangle ) \\
(1/\sqrt{2})(\mid \uparrow \downarrow \rangle \nonumber
-\mid \downarrow \uparrow \rangle )
\end{eqnarray}
To get the last two mutually conjugate measurements
in our set of five, we rotate the second of the two spins by 
$120^\circ$ around the vector $(1,1,1)$---in one direction
or the other---and {\em then}
perform the Bell measurement.  The rotation cyclically 
permutes the $x$, $y$, and $z$ axes.  If the
rotation is in the direction $z \rightarrow y \rightarrow
x \rightarrow z$, I will call the resulting measurement
the ``belle'' measurement, and if the rotation is in
the opposite direction, we get the ``beau'' measurement.
Thus our five mutually conjugate measurements are $XX$,
$YY$, $ZZ$, belle, and beau. 

Of course one does not have to restrict one's attention
to {\em orthogonal} measurements.  A single generalized
measurement (POVM) with the right number of outcomes, performed
on many members of the given ensemble, could be used to 
determine the state just as efficiently as our conjugate-measurement
scheme.  I am focusing on the conjugate-measurement scheme
mainly because it will lead us to the phase space picture that
I want to develop.\footnote{Actually it is conceivable that
the same phase space picture can be used to find 
an optimal generalized measurement for state 
determination---optimal,
that is, if each pair of qubits is to be measured
independently of the other pairs---since
the number of points of phase space is the same
as the number of outcomes a generalized measurement
would need to have in order to be good for state
determination.  But this is a problem for future research.}
But in fact mutually conjugate measurements are also interesting
in another context,
quantum cryptography.  From the earliest work on that
subject conjugate measurements for qubits have played a special
role \cite{Wiesner,Bennett}---Steven Wiesner's original paper is titled
``Conjugate Coding''---and more recently conjugate measurements
in higher-dimensional state spaces have likewise been applied
to cryptography \cite{crypto}.  They have also been used in
a more general analysis of the principle that underlies
quantum cryptography, namely, the trade-off between gaining
information and preserving the state \cite{Barnum}.

It is thus of interest to determine how many mutually 
conjugate measurements one can find in a general $N$-dimensional state
space.  The question is really about mutually conjugate
{\em bases}, since the eigen{\em{values}} that one might
assign to the outcomes of a measurement are not relevant
either for state determination or for cryptography.
Two orthonormal bases are mutually conjugate if, given 
any vector $|v\rangle$ from one of the bases and any vector
$|w\rangle$ from another one, the magnitude of the inner
product, $|\langle v|w\rangle |$, has a fixed value independent
of the choice of vectors---in
fact this value must be $1/\sqrt{N}$ in order for the 
vectors to be normalized.
The following facts represent our current state of knowledge
about the problem of finding
mutually conjugate bases.\footnote{At least, these facts
represent the current state of knowledge of physicists 
whose work
on the problem I am familiar with.  
There is always the possibility that somewhere in
the mathematics literature one might find a paper that holds
more answers.}  (1) In a complex vector space of $N$ dimensions,
there can exist at most $N+1$ mutually conjugate 
bases \cite{Delsarte,Ivanovic}. 
This is interesting because for state determination, the
minimum number of orthogonal measurements needed is also
$N+1$, this being the ratio of the number of parameters 
required to specify a state and the number
of independent probabilities one obtains from each measurement:
$(N^2-1)/(N-1)=N+1$.
(2) If $N$ is a power of a prime, then there do exist $N+1$ mutually 
conjugate bases.  
Moreover, a number of
methods have been devised for constructing such bases in this
case \cite{Fields,Calderbank,Bandy,Lawrence,Chaturvedi}.  
(3) For every $N$ that is not a power of a prime, it is not known
whether $N+1$ such bases exist.  This is true even for $N=6$.

One interesting feature of the discrete phase space construction
described in this
paper is that it provides a novel way a generating a complete set
of mutually conjugate bases in $N$ dimensions when $N$ is a power
of a prime.  Before we get to that construction, though, let us
review briefly the best known 
phase-space representation of quantum states of
continuous systems, the Wigner function.

\section{The Wigner function and quantum tomography}

Let $\rho$ be the density matrix of a quantum particle moving
in one dimension.  The Wigner function $W(q,p)$ is an alternative representation
of the quantum state of such a particle \cite{Wigner}.  
It is defined by
\begin{equation}
W(q,p) = \frac{1}{\pi\hbar}\int \langle q-x|\rho|q+x\rangle
\,\exp(2ipx/\hbar)\, dx.
\label{Wigner}
\end{equation}
Here $q$ and $p$ are the particle's position and momentum;
so the Wigner function is a real function on ordinary 
phase space.  The integral of the Wigner function over
all of phase space is unity, as it would be for a probability
distribution, but the Wigner function is not a probability
distribution: it can take negative values.

There is, however, an interesting respect in which the Wigner
function does act like a probability distribution.
Consider any two parallel
lines in phase space, described by the equations $aq+bp = c$
and $aq+bp = c'$, where $a$, $b$, $c$, and $c'$ are real
constants.  One can show that the integral of the Wigner function over the
infinite strip of phase space between these two lines is
equal to the probability that the {\em operator} $a\hat{q}
+b\hat{p}$ will be found to take a value between $c$ and 
$c'$ \cite{Wootters,tomography}.  
In other words, the integral of the Wigner function
over any direction in phase space yields the correct probability
distribution for an operator associated with that direction.
Here a ``direction'' is defined by a complete set of parallel
lines.  Such sets of lines will be important 
for our discrete phase space as well, and we will call them
``striations'' of the phase space.

The Wigner function has many other special properties
\cite{review}, of
which I will mention just one: translational covariance.
Let $\rho$ be an arbitrary state of our one-dimensional
particle, and let $W$ be the corresponding Wigner 
function.  The state $\rho'$ defined by 
$$
\rho' = \exp[i(q_0\hat{p}-p_0\hat{q})/\hbar]\,\rho \, \exp[-i(q_0\hat{p}-p_0\hat{q})/\hbar]
$$
represents the result of translating $\rho$ by a displacement $q_0$
in space and boosting its momentum by an amount $p_0$.  
When the state is displaced in phase space in this way, the
Wigner function follows along as one expects it should: if
$W'$ is the Wigner function associated with $\rho'$, then
$$
W'(q,p) = W(q-q_0,p-p_0).
$$
When we generalize the Wigner function to discrete systems,
we will insist on an analog of this translational covariance.

We can use the relation between the Wigner function and actual
probability distributions as the basis of a method for determining
the quantum state of our one-dimensional particle, assuming that we are given
a large ensemble of such particles all described by the same density
matrix $\rho$.  For each
direction in phase space, that is, for each striation, we perform
on a subensemble the measurement associated with that striation.
Such a measurement can always be represented by an operator of 
the form $a\hat{q}+ b\hat{p}$, and the observed probability
distribution for that operator gives us the integral of $W$ along the given
direction.  From the integrals of $W$ over every direction,
it is mathematically possible to reconstruct $W$ itself.  The 
process is closely analogous to medical tomography and is in fact
called quantum tomography \cite{tomography,Leonbook}.  Once we have found the Wigner function,
we have found the particle's state, because the Wigner function is an expression
of that state.  However, if one prefers the
density matrix, one can find it via the inverse of Eq.~(\ref{Wigner}):
$$
\langle q_1|\rho |q_2\rangle = \int W\left(\frac{q_1+q_2}{2},p\right)
\,\exp[ip(q_1-q_2)/\hbar]\, dp.
$$
Of course in real life we cannot literally perform the infinite
number of measurements required by this scheme, but we can estimate
the state by performing a large but finite number of measurements.

On first sight quantum tomography may seem impractical because
one does not know how to measure a general operator of the form
$a\hat{q} + b\hat{p}$.  But there is a special case for which
such measurements are very natural, namely, 
the case of a harmonic
oscillator.  As the quantum state of a harmonic oscillator evolves
in time, its Wigner function simply rotates rigidly around the 
origin of phase space, at least if the relative scale of the
position and momentum axes is chosen in the right way.  Thus to
measure the operator associated with some skew direction, we
can simply allow the system to evolve for the right amount of
time and then measure the position.
Now, a mode of the electromagnetic field can be regarded as
a harmonic oscillator.  So quantum tomography is particularly 
suited to finding the quantum state of such a mode, and indeed,
quantum tomography has mostly been used in quantum optics.  

Notice the similarities between the tomography just described
and the method of state determination for one or two qubits
described in Section~{\ref{state}}.  In both
cases we use a set of orthogonal measurements which are
sufficient for reconstructing all the parameters of the
density matrix.  If fact, it even turns out that the measurements
used in the continuous case are all mutually conjugate:
that is, each eigenstate of the operator
$a\hat{q} + b\hat{p}$ yields a uniform distribution of the
value of $a'\hat{q} + b'\hat{p}$, as long as $(a,b)$ and $(a',b')$ 
define different directions in phase space.  The most glaring
lack of similarity between the two cases is that in the
continuous case the measurements arise naturally from the
phase space description of the particle, whereas in the 
discrete case there is no such description, and the 
measurements are constructed by other means.  The rest of
this paper is motivated by the following questions: Can the
measurements that we used for determining the state of one
or two qubits be obtained from a phase space description
of the system?  And if so, can this construction be generalized
to larger systems?  As we will see, the answer to both questions
is yes.

\section{Phase space for a single qubit}

\begin{figure}[h]
\centering
\includegraphics{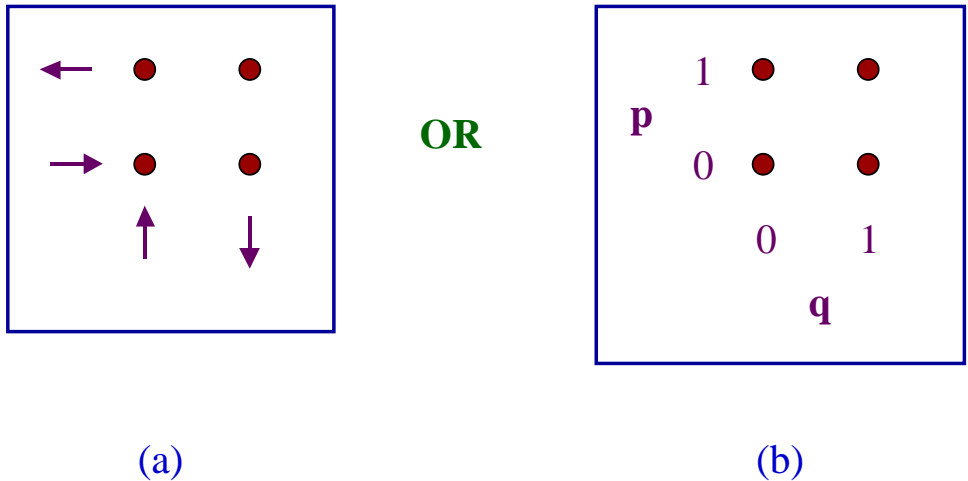}  
\caption{(a) Labeling the phase space with physical states.
(b) Labeling the phase space with abstract variables $q$ and
$p$ taking values in $\mathbb{Z}_2$.}
\label{2x2}
\end{figure}

Let us consider first
the case of a single qubit, imagined as a single spin-1/2
particle.  I will take the horizontal axis of phase space
to represent the $z$ component of spin, which takes the
two values $\uparrow$ and $\downarrow$.  The vertical axis
will represent the $x$ component of spin, its two values
being $\rightarrow$ and $\leftarrow$.  Thus the phase space
consists of exactly four points, as shown in Fig.~\ref{2x2}a.

In order to make sense of the notion of a ``line'' in the
discrete phase space, and the notion of ``parallel lines,''
we want to be able to write down algebraic equations involving
the phase space variables.  So, in addition to associating 
with the axes the physical states shown in Fig.~\ref{2x2}a,
we also want to associate with these axes two
variables $q$ and $p$, analogs
of position and momentum, that take numerical values.  
I will let these numerical values be 0 and 1, interpreted
as elements of the binary field $\mathbb{Z}_2$.  That is,
addition and multiplication of the values of $q$ and $p$
will be mod 2.  This way of labeling the phase space is
shown in Fig.~\ref{2x2}b.  

\begin{figure}[h]
\centering
\includegraphics{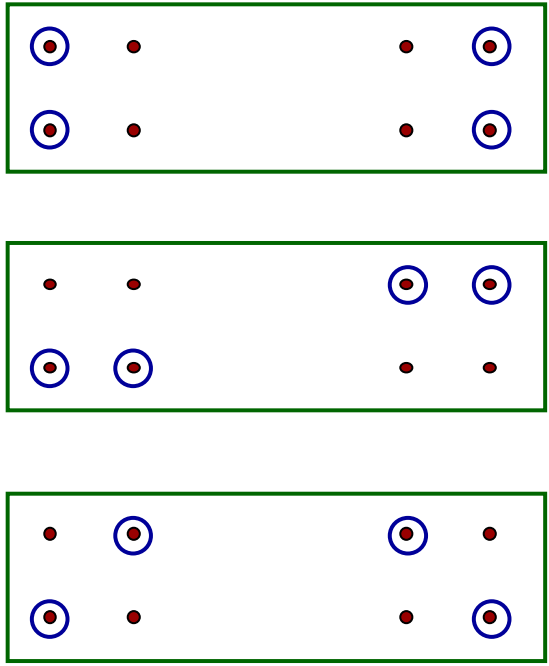}  
\caption{The three striations of the $2\times 2$ phase
space.}
\label{2stri}
\end{figure}

A line in this phase space is the set of points that
satisfies a linear equation, $aq+pb = c$, where $a$, $b$,
and $c$ also take values in $\mathbb{Z}_2$.  For example,
the equation $q+p=0$ defines the line consisting of the
two points $(0,0)$ and $(1,1)$.  It is parallel to the
line defined by $q+p=1$, which consists of the points
$(0,1)$ and $(1,0)$.  In fact there are exactly three
sets of parallel lines in this phase space, that is,
three striations, and these are shown in Fig.~\ref{2stri}.

\noindent As in the continuous case, each striation will be associated
with a measurement, and each line in the striation will be associated with
a particular outcome of the measurement.  Shortly we will 
define a Wigner function on this phase space, which will represent
an arbitrary spin state by four real numbers, one for each
point in phase space.  The Wigner function will 
have the property that its sum over
any line is equal to the probability of the measurement outcome associated
with that line.  

We are thus led to the following question: What measurement---or more properly,
what orthogonal basis---are we to associate with each striation?  
In labeling the axes, we have implicitly associated bases with
the horizontal and vertical striations: the vertical lines 
are associated with the states $\mid\uparrow\rangle$ and $\mid\downarrow\rangle$,
and the horizontal lines are associated with the states
$\mid\rightarrow\rangle$ and $\mid\leftarrow\rangle$.
So it only remains to associate a basis with the diagonal
lines.  The reader is likely to be able to guess
what basis we will assign to these lines, but I want the
structure of the phase space to pick out this basis for us,
as if we could not guess it.  (The construction will be more
impressive in the case of two qubits, where it is harder to
guess the measurements.)

The crucial concept for fixing the remaining basis is
the concept of a {\em translation} in the discrete phase space.  
A translation is simply the addition, mod 2, 
of a vector $(q_0, p_0)$ to each
point in phase space.  Note that, according to our physical
interpretation of the axes, translating by one unit
in the horizontal direction---that is, adding the vector
$(1,0)$---amounts to interchanging
$\uparrow$ and $\downarrow$ while leaving $\rightarrow$
and $\leftarrow$ unchanged.  Physically, these changes 
correspond to a rotation of the spin by $180^\circ$ around
the $x$ axis, which is represented mathematically by the
unitary operator $\sigma_x$, one of the Pauli matrices.  
We therefore associate a unit horizontal translation on phase space with
the operator $\sigma_x$ on state space, and we refer
to this operator as $H$, the horizontal translation 
operator.  (The operator $\sigma_x\exp(i\phi)$,
with an arbitrary phase $\phi$, would work just as well.  For 
definiteness we choose $\phi$ to be zero.)
Similarly a unit vertical translation must be associated
(up to an overall phase factor) with the operator
$\sigma_z$, which we call $V$.  These two unitary operators are analogous to
the operators $\exp(iq_0\hat{p}/\hbar)$ and 
$\exp(-ip_0\hat{q}/\hbar)$, which effect translations
in the continuous phase space.

We will want our Wigner function to be translationally covariant,
like the continuous Wigner function.  For example, given a Wigner
function $W$ that represents some spin state $\rho$, if we change
$\rho$ by applying $H$, we want $W$ to change by a
horizontal translation; that is, we want the values of $W$ to
be swapped in horizontal pairs.  In order to achieve
covariance for this particular translation, that is, 
the horizontal translation, we insist that 
the basis that we associate with the horizontal lines be the
basis of eigenvectors of $H$.  But these eigenvectors
are $\mid\rightarrow\rangle$ and $\mid\leftarrow\rangle$,
which, not surprisingly, we have already decided to associate 
with the horizontal lines.  So translational
covariance does not tell us anything new about the horizontal
lines.  

The requirement of translational covariance does give us something
new, however, when we apply it to the diagonal lines.  In phase
space, the diagonal lines are invariant under a combined vertical
and horizontal translation.  Therefore, the basis we associate 
with the diagonal lines is the basis of eigenvectors of
$VH$.  But $VH = \sigma_z\sigma_x = i\sigma_y$, and the
eigenvectors of this matrix are the states associated with the
spin directions ``into the paper'' and ``out of the paper.''
Notice that we would have arrived at these same states if we had
multiplied our translation operators in the opposite order:
$HV = -i\sigma_y$.  
So these states constitute our third basis, which we associate 
with the diagonal lines.  Thus we started
with two conjugate bases, along the $z$ and $x$ directions, and
our construction
automatically produced a third conjugate basis, along the
$y$ direction.

There remains one ambiguity to clear up before we can define
the Wigner function for a single qubit.  Though our construction
with translation operators tells us what basis to associate with
the diagonal lines, it does not tell us which basis vector to
associate with each line.  There are two choices which are
equally natural.  Let us arbitrarily associate the $+1$ eigenstate
of $\sigma_y$ with the line $\{(0,0),(1,1)\}$; then the $-1$
eigenstate is associated with the line $\{(0,1),(1,0)\}$.
Once this choice is made, the Wigner function of any spin state
is determined by the requirement that its sum over any line is
equal to probability of the measurement outcome associated with
that line.  I will not go into the detailed construction here,
but will simply give examples of Wigner functions of particular
states.    

\bigskip

\begin{center}
\begin{tabular}{c @{\hspace{17mm}} c}
{\bf state} & {\bf Wigner function} \\
 & \\
$\mid\uparrow\rangle$ & $\begin{array}{|c|c|}
\hline
\hbox{\scriptsize $\frac{1}{2}$} & 0 \\
\hline
\hbox{\scriptsize $\frac{1}{2}$} & 0 \\
\hline
\end{array}$ \\
 & \\
$\mid\rightarrow\rangle$ & $\begin{array}{|c|c|}
\hline
0 & 0 \\
\hline
\hbox{\scriptsize $\frac{1}{2}$} & \hbox{\scriptsize $\frac{1}{2}$} \\
\hline
\end{array}$ \\
 & \\
\begin{tabular}{c}
$+1$ eigenstate \\
of $\sigma_y$
\end{tabular} & $\begin{array}{|c|c|}
\hline
0 & \hbox{\scriptsize $\frac{1}{2}$} \\
\hline
\hbox{\scriptsize $\frac{1}{2}$} & 0 \\
\hline
\end{array}$ \\
 & \\
\begin{tabular}{c}
$-1$ eigenstate of\\
$(\sigma_x+\sigma_y+\sigma_z)/\sqrt{3}$
\end{tabular}
& $\begin{array}{|c|c|}
\hline
0.394 & 0.394 \\
\hline
-0.183 & 0.394 \\
\hline
\end{array}$ \\

\end{tabular}
\end{center}

\bigskip

\noindent The negative number in the last example is
the most negative value possible for our one-qubit
Wigner function.  Note that the sums over lines
are legitimate probabilities; for example,
in each case the sum of $W$ over the two points $(0,0)$ and
$(1,1)$---that is, the lower left and the upper right---is the correct probability of finding the particle with
its spin pointing in the positive $y$ direction.  

The essential features of this phase-space description of a single
spin-1/2 particle were proposed independently 
by Feynman \cite{Feynman}
and Wootters \cite{Wootters} in 1987, and a similar though
not identical construction had been worked out a year earlier
by Cohen and Scully \cite{Cohen}.   
So the $2 \times 2$ phase space has been around for a while.  However,
the generalization to two qubits presented in the following
section is new.

\section{Phase space for a pair of qubits}

\begin{figure}[h]
\centering
\includegraphics{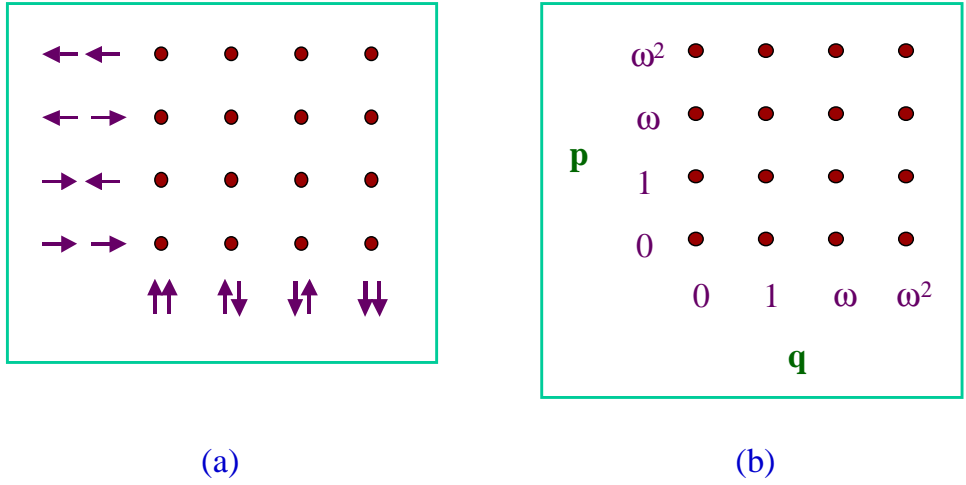}  
\caption{(a) Labeling the $4\times 4$ phase space with 
physical states.  (b) Labeling the same phase space with
the variables $q$ and $p$ which take values in $\mathbb{F}_4$.}
\label{4x4}
\end{figure}

The state space for a pair of qubits has four dimensions;
so I will take the phase space for this system to be
a $4\times 4$ array of points.
Imagining the qubits as spin-1/2 particles,
I will associate the points of the 
horizontal axis of phase space with
the states $\mid\uparrow\uparrow\rangle$, $\mid\uparrow\downarrow\rangle$, 
$\mid\downarrow\uparrow\rangle$, 
and $\mid\downarrow\downarrow\rangle$.  The vertical
axis will represent the conjugate basis consisting of 
$\mid\rightarrow\rightarrow\rangle$, $\mid\rightarrow\leftarrow\rangle$, $\mid\leftarrow\rightarrow\rangle$,
and $\mid\leftarrow\leftarrow\rangle$.  This labeling of the axes is shown
in Fig.~\ref{4x4}a.    

As in the case of a single qubit, we also want to label the axes
with variables $q$ and $p$ that take numerical values.  The choice
of the numerical values is perhaps the most novel feature presented
in this
paper: $q$ and $p$ will take values in the four-element field
$\mathbb{F}_4$, whose elements I will write as 
$\{0,1,\omega,\omega^2 \}$.  We thus get 
the labeling shown in Fig.~\ref{2x2}b.
The arithmetic of $\mathbb{F}_4$ is the only commutative, associative,
and distributive arithmetic on four elements in which both addition
and multiplication have inverses \cite{field}.  This arithmetic is defined by
the following relations:
\begin{eqnarray}
& 1+1 = \omega + \omega = \omega^2 + \omega^2 = 0 \nonumber \\
& 1+\omega = \omega^2 \nonumber \\
& (\omega)(\omega^2) = 1  \nonumber
\end{eqnarray}
Notice that it is not the same as arithmetic mod 4.  In arithmetic
mod 4 there is no multiplicative inverse, because there is no number by which
we can multiply 2 to get 1.  

The arithmetic of the four-element field actually makes some physical
sense for our pair of spin-1/2 particles.  Consider, for example, a
horizontal translation by the field element 1.  This translation
interchanges the first two columns ($0 \leftrightarrow 1$) as well
as the last two columns ($\omega \leftrightarrow \omega^2$).
According to the labeling given in Fig.~\ref{2x2}, 
this permutation corresponds
to interchanging $\uparrow$ and $\downarrow$ for the second particle (while keeping $\rightarrow$ and $\leftarrow$ unchanged), and leaving
the first particle entirely unaffected.  Physically this corresponds
to rotating the second particle by $180^\circ$ around the $x$ axis.
All the other translations
on this phase space, defined by adding other 
vectors under the addition
rules of $\mathbb{F}_4$, similarly have simple physical interpretations. 
From these interpretations we can directly write down unitary
translation operators corresponding to the phase-space translations.
For example, the translation mentioned above---a horizontal translation
by the field element 1---is associated with the unitary operator
$I\otimes \sigma_x$.  We call this operator $H_1$, the subscript
indicating the field element by which we are translating.  For this
two-qubit case there are four basic translation operators, which I
list here.
\begin{eqnarray}
H_1 = I \otimes \sigma_x \nonumber \\
H_\omega = \sigma_x \otimes I \label{trans} \\
V_1 = I \otimes \sigma_z \nonumber \\
V_\omega = \sigma_z \otimes I \nonumber
\end{eqnarray}
All other translations can be obtained as combinations of
these four.  For example, translating by the vector $(1, \omega^2)$
can be decomposed into a horizontal translation by 1, a vertical translation
by $\omega$, and a vertical translation by 1 (since $\omega^2 = 1+\omega$).  
This translation is
therefore associated with the unitary operator $H_1V_\omega 
V_1 = -i\sigma_z \otimes \sigma_y$.  (As before, the order of multiplication
of the $H$'s and $V$'s only affects the overall phase factor, which will
not be significant for any of what follows.)

\begin{figure}[h]
\centering
\includegraphics{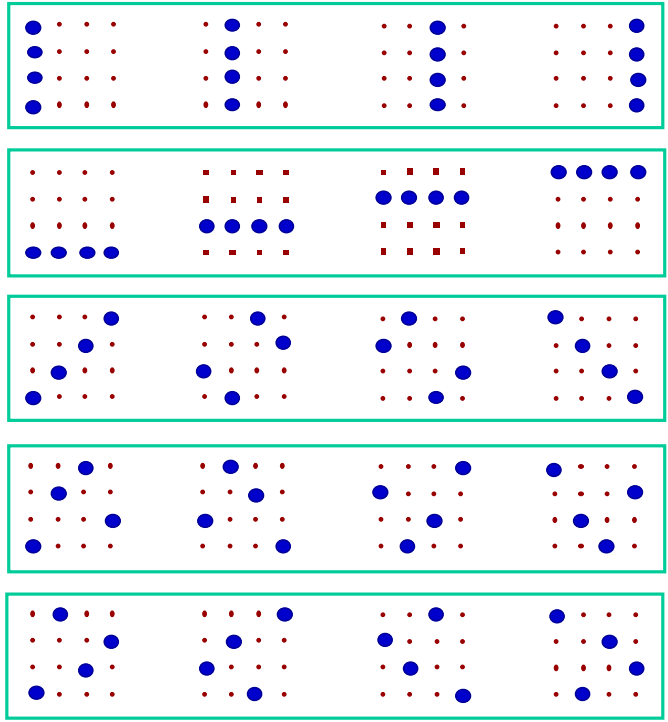}
\caption{The five striations of the $4\times 4$
phase space.}
\label{4stri}
\end{figure}

As in the case of a single qubit, the notion of a {\em striation}
is crucial to our construction.  Again, lines
are defined as the solutions to linear equations, and two lines
are parallel if they can be represented by equations that
differ only in the constant term.  One finds that in the two-qubit
phase space there are exactly five striations, which are shown
in Fig.~\ref{4stri}.  Though the lines may not look like lines,
notice that the usual rules about parallel lines in a plane hold
in this space as well: if two lines are parallel, they have
no point in common, and if two lines are not parallel, they have
exactly one point in common.  These rules follow from the
fact that arithmetic in a field is so well behaved, particularly
that multiplication is invertible.  If we had defined our lines
on the basis of mod 4 arithmetic---this arithmetic produces
 ``wrap-around
lines''---it would have been possible for two distinct lines to
have {\em two} points in common.  

In order to define a Wigner function, we need to associate 
an orthogonal basis with each striation.  To see how this association is 
done, let us consider as an example the fourth striation listed
in Fig.~\ref{4stri}.  Notice that each line of this striation
is invariant under translations by the vectors $(1,\omega)$,
$(\omega,\omega^2)$, and $(\omega^2,1)$.  These translations are associated with
the unitary operators $H_1V_\omega$, $H_\omega V_\omega V_1$, 
and $H_\omega H_1 V_1$, respectively.  Therefore, we want the 
basis vectors that we assign to these lines to be eigenvectors
of all three operators, if that is possible.  One can check that
these three operators commute with each other; so it is indeed
possible
to find simultaneous eigenvectors.  Moreover this criterion
picks out a {\em unique} basis, which I write out here
in the standard representation in which $\sigma_z 
= \mtx{cc}{1 & 0 \\ 0 & -1}$ and $\sigma_x =
\mtx{cc}{0 & 1 \\ 1 & 0}$.   
\begin{equation}
\frac{1}{2}\mtx{c}{1\\1\\i\\-i} \hspace{6mm}
\frac{1}{2}\mtx{c}{1\\1\\-i\\i} \hspace{6mm} \label{basis}
\frac{1}{2}\mtx{c}{1\\-1\\i\\i} \hspace{6mm}
\frac{1}{2}\mtx{c}{1\\-1\\-i\\-i} 
\end{equation} 
One can verify that this is the ``belle'' basis described
in Section~\ref{state}.  In the same way, one finds
that the bases associated with the other four striations
are precisely the other mutually conjugate bases listed
in Section~\ref{state}.  In the order in which the striations
are listed in Fig.~\ref{4stri}, the corresponding bases are
$ZZ$, $XX$, $YY$, belle, and beau.  So the measurements
we imagined using for state determination {\em do} come
from a phase space, just as they do in the continuous
case.  
  
Let us recapitulate the steps by which we obtain an 
orthogonal basis from each striation.  (1) Two conjugate bases are chosen at the beginning
to be associated with the vertical and horizontal striations.
(2) From this assignment, one derives a set of unitary
operators on state space that correspond to
the translations of phase space.  (3) Each striation
defines a set of phase-space translations that preserve
the lines of that striation.  (4) The simultaneous
eigenvectors of the corresponding {\em unitary} operators
constitute the basis that we associate with the given striation.

This procedure raises a number of questions, mostly having to
do with its potential generalization to other dimensions.  
Will it always happen that all the
translation operators associated with
a given striation commute with each other?  Will the number
of striations always be equal to the number of 
orthogonal measurements one needs for state determination?
And if every striation {\em does} generate a definite
orthogonal basis, are the bases associated with different
striations guaranteed to be conjugate?   I will address these 
questions in the following section.  For now, I would like
to return to the definition of the Wigner function for
a pair of qubits.  

Even though we now have a definite correspondence between 
striations of the phase space and bases for the state space,
we have not yet specified which {\em vector} in each basis
goes with each {\em line} of the corresponding striation.  
As in the single-qubit case, there is no unique way of 
choosing this assignment.  However, the choices are not completely
arbitrary.  Consider, for example, the first line
of the fourth striation shown in Fig.~\ref{4stri}.  It
consists of the points $(0,0)$, $(1,\omega)$, $(\omega,
\omega^2)$, and $(\omega^2, 1)$.  To this line, we can
assign any of the four ``belle'' 
basis vectors listed in Eq.~(\ref{basis}).
But once we choose one of these vectors,
there is no further choice involving that particular
striation and that particular basis.  This is because
the other lines of that striation can be obtained by
translating the first one.  We can therefore use the following
rule: if line $\lambda_2$ is obtained from line $\lambda_1$
by a translation $T$, then the basis vector we assign to
$\lambda_2$ should be obtained from the basis vector
assigned to $\lambda_1$ by the unitary operator associated
with $T$.  Indeed, one can show that this rule must be
followed if the resulting Wigner function is to be
translationally covariant \cite{Gibbons}.  So overall, there is some
arbitrariness in the assignment of basis vectors to 
lines, but not as much arbitrariness as one might have expected.

Once every line in phase space has been assigned a state
vector from the appropriate basis, the definition
of the Wigner function $W$ is determined:
it is determined by the requirement
that the sum of $W$ over any line in phase space is
equal to the probability of the measurement outcome
associated with that line.  In order to get a definite
Wigner function, I will
choose the following correspondence between lines
and state vectors.  

\bigskip

\begin{center}
\begin{tabular}{c @{\hspace{1cm}} c @{\hspace{1cm}} c}
{\bf line} & & {\bf state} \\
 & & \\
$\begin{array}{cccc}
\circ & \circ & \circ & \bullet \\
\circ & \circ & \bullet & \circ \\
\circ & \bullet & \circ & \circ \\
\bullet & \circ & \circ & \circ 
\end{array}$ & $\longleftrightarrow$ & $\frac{1}{2}\mtx{c}
{1 \\ -i \\ i \\ 1}$ \\
 & & \\
$\begin{array}{cccc}
\circ & \circ & \bullet & \circ \\
\circ & \bullet & \circ & \circ \\
\circ & \circ & \circ & \bullet \\
\bullet & \circ & \circ & \circ 
\end{array}$ & $\longleftrightarrow$ & $\frac{1}{2}
\mtx{c}{1 \\ 1 \\ i \\ -i}$ \\
 & & \\
$\begin{array}{cccc}
\circ & \bullet & \circ & \circ \\
\circ & \circ & \circ & \bullet \\
\circ & \circ & \bullet & \circ \\
\bullet & \circ & \circ & \circ 
\end{array}$
& $\longleftrightarrow$ & $\frac{1}{2}
\mtx{c}{1 \\ -i \\ 1 \\ i}$

\end{tabular}
\end{center}

\bigskip

\noindent These three choices fix the assignments of
vectors to lines for the last three bases---$YY$, belle,
and beau---and therefore fix the definition of the 
Wigner function. 

With these choices, the
Wigner function can be worked out 
for any two-qubit state \cite{Gibbons}, some examples of which
are shown below.\footnote{An algorithm for generating the
Wigner function of any two-qubit state---given the 
choices specified in the text---is the following.
Let $\alpha = (q,p)$ be any point in phase space.  The
Wigner function evaluated at $\alpha$ will be
$W_\alpha = (1/4)\hbox{tr}\,\rho A_\alpha$, where
$\rho$ is the density matrix and $A_\alpha$ is 
a matrix associated with the point $\alpha$.  
The $A$-matrix associated with the origin is
$A_{(0,0)} = \mtx{cc}{1 & (1-i)/2 \\ (1+i)/2 & 0}
\otimes \mtx{cc}{1 & (1+i)/2 \\ (1-i)/2 & 0}$.
Any other $A_\alpha$ can be obtained from
$A_{(0,0)}$ via the translation operators:
$A_\alpha = U_\alpha A_{(0,0)}U_\alpha^\dag$, 
where $U_\alpha$ is the
unitary operator corresponding to a translation
by the vector $\alpha$, as given in Eq.~(\ref{trans}).}

\bigskip

\begin{center}
\begin{tabular}{c @{\hspace{17mm}} c}
{\bf state} & {\bf Wigner function} \\
 & \\
$\mid\uparrow\uparrow\rangle$ & 
$\begin{array}{|c|c|c|c|}
\hline
\hbox{\scriptsize $\frac{1}{4}$} & 0 & 0 & 0 \\
\hline
\hbox{\scriptsize $\frac{1}{4}$} & 0 & 0 & 0 \\
\hline
\hbox{\scriptsize $\frac{1}{4}$} & 0 & 0 & 0 \\
\hline
\hbox{\scriptsize $\frac{1}{4}$} & 0 & 0 & 0 \\
\hline
\end{array}$ \\
 & \\
$\mid\uparrow\rightarrow\rangle$ & $\begin{array}{|c|c|c|c|}
\hline
0 & 0 & 0 & 0 \\
\hline
\hbox{\scriptsize $\frac{1}{4}$} & \hbox{\scriptsize $\frac{1}{4}$} & 0 & 0 \\
\hline
0 & 0 & 0 & 0 \\
\hline
\hbox{\scriptsize $\frac{1}{4}$} & \hbox{\scriptsize $\frac{1}{4}$} & 0 & 0 \\
\hline
\end{array}$ \\
 & \\
$\frac{1}{\sqrt{2}}(\mid\uparrow\downarrow\rangle
- \mid\downarrow\uparrow\rangle)$ & 
$\begin{array}{|c|c|c|c|}
\hline
0 & 0 & 0 & 0 \\
\hline
0 & \hbox{\scriptsize $\frac{1}{4}$} & \hbox{\scriptsize $\frac{1}{4}$} & 0 \\
\hline
0 & \hbox{\scriptsize $\frac{1}{4}$} & \hbox{\scriptsize $\frac{1}{4}$} & 0 \\
\hline
0 & 0 & 0 & 0 \\
\hline
\end{array}$ 
\end{tabular}
\end{center}

\bigskip
\noindent Again one can check that the sums over lines
make sense.  For example, the sum over the first vertical
line in each case is the probability of finding the pair
in the state $\mid\uparrow\uparrow\rangle$.

To return to the problem of state determination, we see
that tomography for a two-qubit system is very similar
to tomography for a continuous system.  Each ``direction''
in phase space, as defined by a striation, is associated
with a measurement, and the collection of measurements
obtained in this way is just sufficient to determine the
state of the system.  It is interesting that, in a paper
on reconstructing the state of discrete quantum systems,
Asplund and Bj\"ork discuss the use of mutually conjugate
bases and refer to these bases as being like different
directions in phase space \cite{Asplund}.  
The above construction based on the four-element field 
shows, at least for two qubits, 
that this analogy is indeed quite apt.

\section{Generalization to other dimensions}

To what extent can the above 
phase-space representations be generalized to state
spaces of other dimensions?  First, it is 
crucial to our construction that the axes be labeled
by the elements of a {\em field}, with its invertible
addition and multiplication.  Now, it is a fact that
there exists a field with $N$ elements if and only
if $N$ is a power of a prime, and in that case, there
is essentially only one field possible \cite{field}.  So our
construction does not apply directly to every quantum system,
but it does apply to a system of $n$ qubits, since the
dimension then is $N = 2^n$, which is a power
of a prime.  When the phase space axes are labeled
by the $N$-element field, it is not hard to show that the number
of striations is $N+1$, exactly the number of orthogonal
bases needed for state determination.  

Moreover, whenever the dimension is a power of a prime,
it turns out that there is a systematic way of labeling
the axes with field elements so that the above construction
always works.  That is, the translation operators
associated with a given striation commute with each other
and define a unique basis of eigenstates, and the bases
thereby derived are guaranteed to be mutually 
conjugate \cite{Gibbons}.
Thus this phase-space
construction provides a new method of 
generating a complete set of mutually conjugate
bases whenever the dimension $N$ is a power of
a prime.  It is closely related to, and indeed
is inspired by, the methods
of Bandyopadhyay {\em et al} \cite{Bandy} and Lawrence {\em
et al} \cite{Lawrence}.  For example, our 
translation operators appear in
these earlier constructions, though not as translation
operators {\em per se}.  
Our method is
distinguished from these others in that
it is based on phase space and is explicitly geometrical.

\section{Discussion}

The phase-space representation presented in this
paper follows many other papers on discrete phase
spaces.  As I have said earlier, Feynman proposed
the $2\times 2$ phase space we are using for a single spin-1/2
particle \cite{Feynman}.  He was interested in the concept of negative
probability and asked whether some or all of the mysteries
of quantum mechanics could be rendered more intelligible if
we could make sense of negative probabilities.  
Discrete phase spaces for higher-dimensional quantum
systems have been proposed by a number of authors.  In
the formulations of Cohendet {\em et al} \cite{Cohendet}, Galetti and de Toledo Piza \cite{Galetti}, 
Leonhardt \cite{Leonhardt}, and Wootters \cite{Wootters},
one sees manifestations of certain number-theoretic issues 
that arise when 
one tries to generalize the Wigner
function to discrete systems.  The work of Cohendet {\em et al},
for example,
applies only to systems with an odd-dimensional state
space.  My own earlier work, as well as that of
Galetti and de Toledo Piza, applies most naturally to systems
for which the dimension $N$ of the state space is prime,
though one can also apply it to any composite value
of $N$  
by treating
each prime factor separately.
Leonhardt, who introduced a systematic, phase-space
approach to finite-state tomography, finds problems with 
even-dimensional state
spaces that he avoids 
by making his discrete phase space a $2N \times 2N$
grid when $N$ is even.  Other approaches use
an $N \times N$ phase space for arbitrary $N$ 
but do not insist on any special properties associated with
striations other than those defined by the two axes \cite{Luis,
Takami}.  
A discrete Wigner function adapted particularly 
to quantum optics was introduced by 
Vaccaro and Pegg \cite{Pegg}
(see also the review by 
Miranowicz {\em et al} \cite{Miranowicz}), and a
Wigner function applicable whenever
the configuration space is a finite group of odd order
has been developed by Mukunda {\em et al} \cite{Mukunda}.  
Recently various authors have 
used both the Wigner function of Ref.~\cite{Wootters}
for prime $N$ \cite{Koniorczyk}, and Leonhardt's $2N \times 2N$
formulation generalized to all $N$ \cite{Paz}, to
analyze quantum information processes such as
Grover's search algorithm and teleportation.  

The work I have described here follows most naturally from
Ref.~\cite{Wootters} and is essentially a generalization
of that paper from the primes to the powers of primes.  
This
new work is also more systematic than Ref.~\cite{Wootters}
in that it brings out the 
choices one needs to make in defining a discrete Wigner function.
Further details about these choices, and about how many
truly distinct definitions are possible, are spelled out
in Ref.~\cite{Gibbons}.

As the authors of Refs.~\cite{Koniorczyk} and \cite{Paz} have 
already shown, 
there is some value in using
phase space to visualize the effects of quantum information
processing.  I am hoping that the Wigner function
described here will have its own advantages in this respect.
Another possible application is in the foundations of 
quantum mechanics.  Hardy \cite{Hardy} and Spekkens \cite{Spekkens} have recently
proposed toy models of quantum mechanics that facilitate
the study of certain foundational issues.  Our discrete
Wigner function appears to provide a natural framework in which
to express these toy models and relate them to 
standard quantum mechanics.  For example, in both
Hardy's and Spekkens' models, a ``toybit'' has exactly
four underlying ontic states, which could be taken to correspond
to the four points of our one-qubit phase space.  Moreover these
models allow
exactly six pure epistemic
states, which correspond to the six 
one-qubit Wigner
functions in which two of the four values are zero and the other
two are 1/2.    

For many purposes, it is good to have a way of picturing 
quantum states.  Discrete Wigner functions allow 
such picturing in that they require us to imagine
only three dimensions:
two for phase space itself and one for the value of $W$.  
I have to admit that as a picture, a discrete Wigner
function will never be as graceful as, say, a Bennett 
photograph, but it may give us a new perspective
on some of the remarkable things that Charlie and others
have shown can be done with qubits.

\end{document}